\documentstyle[12pt]{article}

\title{SPINNING PARTICLE AS A\\
SUPER BLACK HOLE}
\author{A.Burinskii\\
Gravity Research Group, IBRAE Academy of Sciences\\
B.Tulskaya 52,Moscow 113191,Russia, e-mail:grg@ibrae.ac.ru}

\begin{document}

\maketitle

\begin{abstract}
  A natural combined model of the Kerr spinning particle and
superparticle is obtained leading to a non-trivial super black hole
solution.

        By analogue with complex structure of the Kerr solution
we perform a supershift on the Kerr geometry, and then select a
"body"-submanifold of superspace that yields a non-trivial
supergeneralization of the Kerr metric with a nonlinear realization
of (2,0)-supersymmetry.

    For the known parameters of spinning particles this "black hole"
is to be in a specific state without horizons and very far from extreme.

    The naked Kerr singular ring has to be hidden inside a rotating
superconducting disk, built of a supermultiplet of matter fields.

Stringy wave excitations of the singular ring (traveling waves) yield
an extra axial singular line modulated by de Broglie periodicity.

\end{abstract}

\medskip
{E-print: hep-th/9802110}, talk given at the MG8 Meeting in Jerusalem
(Israel, June 1997), at the Hyderabad Symposium (India, December 1997),
and at the GR15 Conference (Pune, India, December 1997), brief version
of the paper Phys.Rev.{\bf D57},2392(1998). Submitted to the Proceedings
of MG8 Meeting.
\def\NCA{\em Nuovo Cimento}
\def\NIMA{{\em Nucl. Instrum. Methods} A}
\def\NPB{{\em Nucl. Phys.} B}
\def\PLB{{\em Phys. Lett.}  B}
\def\PRL{\em Phys. Rev. Lett.}
\def\PRD{{\em Phys. Rev.} D}
\section{Introduction}
 The Kerr-Newman BH-solution has paid an attention as a model
of spinning particle since 1968 (Carter, Israel, Burinskii, L\'opez).
Recently, an important role of the Kerr black hole in string
theory has been also obtained and there has appeared a point of
view that black holes should be treated as fundamental string
states \cite{Sen1}, \cite{Dab}, or as elementary particles
\cite{Sen2} - \cite{Bur2}.

On the other hand, the models of spinning particles based
on anticommuting parameters are well known.

The aim of this paper is to discuss one very natural way to combine the
Kerr spinning particle and superparticle models. The resulting background
has a metric of a non-trivial super black hole with a nonlinear realization
of supersymmetry \cite{Bur2}.
For the known parameters of spinning particles the horizons disappear, and
naked singularity has to be covered by a disk-like source built of
a multiplet of superfields.

\section{Complex shift method}
 Starting point is a "complex shift method" allowing to generate
spinning solutions from spherical symmetrical ones. In complex
representation (initiated by Newman, 1974) Kerr solution can be
obtained from Schwarzschild one by a TRIVIAL complex SHIFT
$(x,y,z)\rightarrow (x,y,z+ia) $.

The Kerr metric  $ g_{ik} = \eta
_{ik} + 2 h k_{i}k_{k} $ is defined by the principal null (P.N.)
congruence $k^i(x)$, and can be represented as a retarded-time
field generated by  a "complex point source" which propagates
in  complex Minkowski space $CM^4$
along a complex "world line"
$x^{i}_{o}(\tau),\quad (i=0,1,2,3),$
 parametrized by a complex time parameter
$\tau = t+i\sigma  = x^{o}_{o}(\tau) .$

The non-trivial twisting structure of P.N. congruence appears on a REAL
SLICE as a consequence of nonlinear (light cone) constraints
$(x - x_0(\tau))^2=0, $ where $x_0 (\tau)$ is position of a
mysterious "complex source", and $x$ are points of the real submanifold.

In the gauge $x_0^0 =\tau$
this equation may be split  as a complex retarded-time equation
\begin{equation}t-\tau =  \tilde r = - (x_i-x_{0i}){\dot x}_0^i,
\label{split}\end{equation}
 where $\tilde r=r+ia \cos\theta$ is a compl`ex radial distance.

The complex light cone is split  into
two families of null planes: "right" $( \psi _{R}$ =const; $\bar{\psi
}_{L}$ -var.) and "left"$ ( \bar{\psi }_{L}$ =const; $\psi _{R}$ -var.).

The rays of the principal null congruence $K$ of the Kerr geometry  are
the tracks of these complex null planes (right or left) on the real  slice
of Minkowski space.
\section{Body-slice, and geometry generated by superworldline}
 Now we would like to generalize this complex retarded-time construction
to the case of complex "supersource" propagating along a super-world-line.
By analogy with above construction we consider a TRIVIAL SUPERSHIFT
leading to a complex "supersource"

\begin{equation} X_0^i (\tau) = x_0^i(\tau) - i \theta\sigma^i \bar \zeta
 + i \zeta\sigma^i \bar \theta. \label{SWL}
\end{equation}
Instead of the nonlinear real slice conditions we introduce now the
"BODY-SLICE"  constraints
\begin{equation}(x_i - X_{0i}) (x^i - X_{0}^i) = 0, \label {SBS}
\end{equation}
where $ X_{0i}$ are coordinates
of  "supersource", and coordinates $x_i$ belong to a "body" of
superspace, that means a submanifold of superspace where the nilpotent
part of $x_i$ is equal to zero.
Selecting the nilpotent parts of this equation we obtain
 the above real slice condition and the extra Body-slice conditions
\begin{equation} [x^i-x_0^i (\tau)]
( \theta\sigma_i \bar \zeta
 + \zeta\sigma_i \bar \theta); \label{odd1}\end{equation}
\begin{equation}
( \theta\sigma \bar \zeta
 + \zeta\sigma \bar \theta)^2 =0.\label{odd2}\end{equation}
The eq. (~\ref{odd1}) may be rewritten by using the spinor null plane
parameters $\psi, \quad\bar\psi$ in the form
\begin{equation}
  (\theta^\alpha\sigma_{i\alpha\dot\alpha}\bar\zeta^{\dot\alpha}
 - \zeta^\alpha\sigma_{i\alpha\dot\alpha}\bar\theta^{\dot\alpha})
\psi^\beta \sigma^i_{\beta\dot\beta} {\bar\psi}^{\dot\beta}
=0\label{odd4}\end{equation}
which yields
\begin{equation}
\bar\psi \bar\theta =0,\qquad\bar\psi \bar\zeta =0
\label{odd5}\end{equation}
which in turn is a condition of proportionality of the commuting spinors
$\bar\psi(x)$ and anticommuting spinors $ \bar\theta$ and $\bar\zeta$
providing the left  null superplanes to reach B-slice.

Taking into account that ${\bar \psi}^{\dot 2}=Y (x),
\quad{\bar \psi} ^{\dot 1}=1$ we obtain
\begin{equation}
{\bar\theta}^{\dot 2} = Y (x){\bar\theta}^{\dot 1} ,\quad
{\bar\theta}^{\dot \alpha} = {\bar\theta}^{\dot 1} {\bar\psi}^\alpha,\quad
{\bar\zeta}^{\dot 2} = Y (x){\bar\zeta}^{\dot 1} ,\quad
{\bar\zeta}^{\dot \alpha} = {\bar\zeta}^{\dot 1} {\bar\psi}^\alpha.
\label{SRS}
\end{equation}
It also gives that $\bar\theta \bar\theta = \bar\zeta \bar\zeta =0$,
and equation (~\ref{odd2}) is satisfied automatically.

Therefore, the  Body-slice condition fixes a correspondence between the
coordinates  $ \bar\theta, \quad \bar\zeta$ and twistor null planes forming
the Kerr congruence.

	The retarded time eq. (~\ref{split}) takes the form
$ t- T =  \tilde R = \tilde r + \eta, $
where $\tilde R = - (x_i-X_{0i}){\dot X}_0^i $ is a superdistance.
$T=\tau + \eta $ is a supertime containing the nilpotent term
\begin{equation}
\eta = i \theta\sigma^0 \bar \zeta (\tau) -i \zeta\sigma_0\bar \theta.
\label{eta}\end{equation}

\section{Supershift of the Kerr solution}
By performing  supershift to the Kerr solution one can note that the
Kerr solution is a particular solution of supergravity field equations
with vanishing spin-3/2 field. The solution with supersource
(~\ref{SWL}) can be obtained from the Kerr
solution by supershift
\begin{equation}
x^{\prime i}  = x^i - i \theta\sigma^i \bar \zeta
 + i \zeta\sigma^i \bar \theta;
\qquad
\theta^{\prime}=\theta + \zeta ,\quad
{\bar\theta}^{\prime}=\bar\theta + \bar\zeta, \label{SG}
\end{equation}
which is a "trivial" supergauge transformation.
However, the subsequent imposition of B-slice constraints is a nonlinear
operation breaking the original supersymmetry, so the arising spin-3/2
field can not be gauged away. However, there survives a nonlinear
realization of (2,0)- supersymmetry.

Starting from tetrad form of the Kerr solution  $ds^2=e^1 e^2 +e^3 e^4,$
where \begin{equation}
e^1 = d \xi - Y d v; \qquad e^2 = d \bar \xi - \bar Y d v; \label{KS1}
\end{equation}
\begin{equation}
e^3 =du+ \bar Y d \xi  + Y d \bar \xi - Y \bar Y d v;
\label{KS3}
\end{equation}
\begin{equation}
e^4 =  d v - h e^3,\label{KS4}
\end{equation}
and using the coordinate transformations (~\ref{SG}) under constraints
(~\ref{SRS}), and also substitution $\tilde R \rightarrow \tilde r$, one
obtains the following tetrad
\begin{equation}
e^{\prime 1} = e^1 + ( A - C^1 {\bar \theta}^{\dot 1} ) dY,\qquad
e^{\prime 2} = e^2 + Ad\bar Y -B^2 d{\bar \theta}^{\dot 1},
\label{S12}
\end{equation}
\begin{equation}
e^{\prime 3} = e^3 - C^3 {\bar \theta}^{\dot 1}dY,\qquad
e^{\prime 4} = dv + \tilde h e^3 + dA - B^4 d{\bar \theta}^{\dot 1},
\label{S34}
\end{equation}
where $dY={\tilde R}^{-1}(P e^1 -P_{\bar Y} e^3) $
and
\begin{equation}
 A=i\sqrt{2}(\theta^1 {\bar \zeta}^{\dot 1}) ,\qquad
B^a =ie^a_i(\zeta \sigma^i  \bar \psi),\qquad
C^a =ie^a_i(\zeta \sigma^i \partial_Y \bar \psi),
\label{ABC}
\end{equation}
\begin{equation}
 \tilde h= m(Re{\tilde R}^{-1})/P^3
\label{th}.
\end{equation}
This is a metric of a NON-TRIVIAL \cite{Aich} super black hole, that means
it cannot be turned into a solution of the Einstein gravity by a supergauge
transformation. The non-triviality is provided by the B-slice constraints.

	Similar treatment can be carried out for the Kerr-Newman solution
and for the Kerr solution generalized by Sen to low energy string theory
\cite{Bur2}.

For the parameters of spinning particles $\mid a\mid \gg m$, and
this "super black hole" is to be in a specific state without horizons and
very far from extreme.
       The naked Kerr singular ring has to be covered by a rotating
superconducting (and probably also superfluid) disk, built of a
supermultiplet of matter fields.

Stringy excitations of the Kerr singular ring (traveling waves \cite{Bur3})
yield an extra axial singular line modulated by de Broglie periodicity.

\section{Acknowlegments}
    I am thankful to  G. Gibbons, F. Hehl, Y. Ne'eman, J. Wess and
B. Sidharth for useful discussions and interest to this paper.

\end{document}